# Polymer Assisted Synthesis of FeNi Nanoparticles

S. K. Karna[1], S. R. Mishra[1], E. Gunapala[2], I. Dubenko[3], V. Malagareddy[1], G. K. Marasinghe[2], and N. Ali[3]

1. Department of Physics, The University of Memphis, Memphis, TN 38138
2. Department of Physics, University of North Dakota, Grand Forks, ND 58202
3. Department of Physics, Southern Illinois University, Carbondale, IL 62901

**Abstract**

Polymer assisted spherical FeNi nanoparticles were prepared via wet chemical method using hydrazine as a reducing agent and polymers (PVP and PEG) as reducing and stabilizing agent. Structural studies performed using XRD and TEM shows uniform dispersion of fine FeNi nanocrystallites in nanocomposite particles. The size and thermal stability of FeNi nanoparticles prepared under same reaction condition was found to be dependent on the type and the molecular weight of the polymer used. However, the magnetic properties of nanocomposite particles were not influenced by the polymers. The study highlights subtle differences in using polymers during the synthesis of alloyed nanocomposite particles.

**Introduction**

Recently there has been a growing interest in alloy nanoparticles of different shape and size such as FeCo [1], FePt [2], CoPt [3] and AuPd [4] due to the unique properties of alloys over metals. FeNi alloys are of great interest due to a variety of useful magnetic and mechanical properties. For example, FeNi is a soft magnetic material with high saturation magnetization, low covercivity, high magnetic permeability, low thermal expansion [5], and excellent corrosion resistance [6]. The nanoparticle form of FeNi alloys also received great attention because of diverse practical applications such as catalysis and medicine (e.g. hyperthermia application). In order to further increase the applicability of alloyed magnetic nanoparticles in bioindustry, it is necessary to modify surfaces of these nanoparticles which provide easy template for biomolecule attachment. Often these structures are classified as nanocomposites. Nanocomposite materials offer distinct of multifunctional properties which could be optimized for a particular application. Metal-polymer nanocomposites exhibit excellent physical, chemical, and mechanical properties that single phase materials does not have [7,8,9].

In the present work we report synthesis of FeNi-polymer (PVP and PEG) nanocomposites using chemical reduction method. Recently polymeric materials have been employed as particle stabilizers in the chemical synthesis of metal colloids and the roles of polymers are generally documented as steric stabilizer or capping agent [10,11]. The control of reaction kinetics using polymers allows simple and versatile route to the synthesis of metal nanocrystals with well defined shapes and size [12]. The polymers interact with the small particles, preventing their agglomeration and precipitation [13,14]. A number of metal-polymer nanocomposites with metals such as Ag, Au, Pd, Pt [15], Ni [16], Cu [17] and other noble metal-polymer have been studied. However, the report for the nanocomposites of magnetic metal alloys with polymer is limited [18,19,20].

Thus, an attempt is made to synthesize monodispersed FeNi nanocomposite particles using PVP and PEG as stabilizing and reducing agent. The aim of the study is two folds viz. to study (1) the role of PVP and PEG in the synthesis of nanocomposite particles, and (2) the thermal and magnetic properties of as prepared nanocomposite particles. It was found that size but not the magnetic properties of FeNi-polymer nanocomposite particles are affected by the type and the molecular weight (MW) of the polymers used. Thus, this work presents an approach for the synthesis of nanocomposite particles with uniform magnetic properties which is independent of the size of the particles.

**Experimental**

**Synthesis**

All chemicals were of analytical grads and were used without further purification. Two sets of samples were prepared by using four different MW of Polyethylene glycol (PEG, MW 200,1000, 8000, and 20,000) and Polyvinyl pyrrolidone (PVP, MW 10,000, 29,000, 40,000 and 55,000). The samples were synthesized by chemical phase reduction method [21]. In sample preparation we mixed 0.1396 g iron (II) sulfate heptahydrate ($FeSO_4.7H_2O$) and 0.1492 g nickel (II) sulfate hexahydrate ($NiSO_4.6H_2O$) in the in 10 ml of water. The solution containing 0.5 ml of cyclohexane and 1.3601 g polymer (either PEG or PVP) were added into the solution at argon atmosphere to forms a colloidal solution. After being ultrasonicating the mixture for 80 minutes at room temperature, the mixture was then heated to 80$^o$C for 80 minutes, the solution composed of 5 ml of hydrazine hydrate ($N_2H_4.H_2O$) with 1.0352 g of sodium hydroxide (NaOH) was added into the heated solution to enhance the reducing power. A rapid reaction took place at that time with black fume and solution turned black in color from greenish one. The solution was aged for 20 minutes and then had been washed several times with water and ethanol by centrifugation to remove extra polymer residues from the sample. The sample were dried at about 80°C and stored in a glass vial.

**Characterization**

The phase identification of samples was performed using Bruker D8 Advanced X-ray diffractometer with Cu-K$_\alpha$ radiation. The patterns were matched with ICCD database. The nanocrystallite size, d, was determined using the Scherrer's equation, $d = 0.89\lambda/\beta\sin\theta$ with the wavelength $\lambda = 1.54056$ Å, the peak width $\beta$, and the Bragg angle $\theta$. Size and shape characterization of samples was carried out using transmission electron microscopy (TEM) JEOL's JEM1200EX II TEM. TEM samples were prepared by dropping a dilute sample of particles onto a carbon coated copper grid and viewing the specimen once the solution adequately dried on the grid. Thermal studies TGA/DTA have been done on pure polymers as well as FeNi-nanocomposites using the Perkin Elmer Diamond thermal analysis system. Magnetic data were taken using SQUID as a function of temperature in field cooled (FC) and zero-field cooled (ZFC) conditions. Sample preparation for the SQUID was carried out by dropping concentrated solutions of the nanoparticles onto the sample holder allowing it to dry under ambient conditions. Sample mass consists of contributions from both the nanoparticles and polymer.

**Results and Discussion**

The phase composition of as-synthesized powders was characterized by XRD, and the XRD patterns of samples are shown in Fig. 1. Three characteristic peaks for FeNi$_3$ (2$\theta$ = 44.3°, 51.5°, 75.9°), corresponding to Miller indices (1 1 1), (2 0 0), (2 2 0), were observed. The absence of XRD peaks, characteristics of α-Fe (i.e. at 2$\theta$ of 65.2) and (FCC)-Ni (i.e. at 2$\theta$ of 44.5°, 51.8° and 76.4°), indicates that no iron or nickel was formed. In addition, no nickel oxides can be detected in the XRD pattern. Besides XRD peaks of (FCC)-FeNi$_3$, Fe$_3$O$_4$ phase can be observed also, which indicates Fe$^{2+}$ was not reduced into Fe completely. The presence of iron oxide in the sample is thus expected from the oxidation of unreacted iron ions in the sample. The iron oxide peaks are more pronounced in FeNi-PEG samples than in FeNi-PVP samples. On applying Scherrer's equation to (111) reflection [22], the size of FeNi nanocrystallite were calculated to be

~ 11 nm for all nanocomposite samples. The size of FeNi nanocrystallite was found to be independent of the type and the MW of the polymers used. These nanocrystallites are found to be dispersed in the nanocomposite matrix as revealed by the TEM images.

TEM images of FeNi-PEG and FeNi-PVP nanocomposite samples are shown in Fig. 2a and b, respectively. Images show well dispersed spherical FeNi-nanocomposite particles prepared using PEG and PVP polymers. FeNi nanocrystallites are found to be dispersed in the composite matrix. Actually, sub-µm sized FeNi nanocomposite particles are secondary particles formed by the aggregation of nm sized primary particles. It is still not well understood why, in certain similar cases, the agglomeration of primary particles does not proceed in an anarchic way but instead leads to monodisperse secondary particles. The effect of MW and the type of polymer is evident from these images.

The mean size of FeNi-nanocomposite particles as a function of MW is shown in Fig. 3. It is interesting to note that the FeNi-PEG particle size increases with the increase in the MW of PEG. A similar increase in the size has been reported for the case of Au-PEG [23] and Fe-PEG metal-polymer composite particles [24]. On the other, FeNi-PVP nanocomposite particle size decreases with the increase in PVP MW. Again similar trend has been reported for Pt and Ag-PVP [25] nanoparticles [26]. The effect of MW on the particle growth is expected based on previous studies of metal particles in the presence of surfactants where surfactants with similar binding ability but with larger size lead to slow particle growth. The difference in observed particle size dependence on the type of polymer used could be explained on the basis of reaction rate. Faster reaction rate leads to the growth of nanocomposite particles while slower leads to decrease in the size of nanocomposite particles. In their study on growth kinetics of Ag in PEG, Longenberger and Mills [14] reported that the growth of nanoparticles in the presence of PEG takes place by metal ion binding to crown ether molecules. Upon oxidation of oxyethelene group, pseudocrown ether structure is disrupted, which results in migration of metal ions followed by coalescence of metal atoms with formation of metal cluster. Since with the increase in MW of PEG, cavities are more abundant, the particle formation process is faster with increasing number of cavities. On the other hand the low reaction rate proceeds for polymers that do not form pseducrown either structure, such as PVP. Furthermore, strong adsorption of PVP on the nanocrystallite surface is an obstacle to diffusion of metal ions and may decrease the nanoparticle grain growth [27]. It is also argued that after upon oxidation, PEG become more hydrophilic while PVP hydrophobic. Which could also be the factor leading to the observed variation in the size of FeNi-polymer nanocomposites.

The inset of Fig.3 shows the dependence of FeNi nanocrystallite size, as calculated using Scherrer's equation, on the MW of polymers. It is evident from this figure that on an average FeNi nanocrystallite size is independent of the type and MW of the polymers used. Furthermore, increase in the FeNi nanocrystallite size to ~23 nm is observed upon heating the nanocomposite to 500 $^{o}$C. This increase in the crystallite size upon heating is obvious as the disordered nanocrystallites grow upon heating.

Figure 4(I) compares the DTA curves of pure polymers and FeNi-polymer nanocomposites. Pure polymers show endothermic peak corresponding to melting of polymers at around 60 $^{o}$C and a broad exothermic peak corresponding to degradation of the polymers at around 328 and 376 $^{o}$C for PEG and PVP, respectively. The DTA curves of FeNi-polymer nanocomposites differ from the thermal behavior of pure polymers. The FeNi-polymer nanocomposite show two exothermic peaks at around 260 and 487 $^{o}$C. The first peak corresponds to the temperature of degradation of polymers and second peak correspond to the oxidation of nanocomposite. The decrease in the degradation temperature of FeNi-polymer nanocomposite to 256 $^{o}$C, as compared to 328 $^{o}$C of pure polymers, indicate that the polymers in the nanocomposite is different in its chemical nature compared to that in the pure state. A similar trend is observed in Pt-polymer nanocomposite [28], where the author proposes that the interaction of polymers with Pt nanoparticle surfaces lead to formation of new compound of polymers which do not decompose in simple molecules such as carbon dioxide, instead PVP decompose into some new organic molecule which stick to the particles. However, the interpretation of the result needs further investigation. The oxidation of nanocomposites at higher temperature leads to a mark increase in the iron oxide content in the nanocomposite. This is evident from the XRD plot of nanocomposites obtained at 500 $^{o}$C, Fig. 4(I) inset and TGA results, Fig. 4II. Similar oxidation behavior of nanocomposite has been reported by Cushing et al. in their studies of gold coated FeNi$_3$ nanoparticles [29].

The ZFC and FC susceptibility as a function of temperature measured at 1 kOe, Figure 5, shows that the nanocomposite particles are blocked just below the room temperature. The decrease in magnetization at temperature below 50 K indicates ferrimagnetic behavior of the nanocomposites. Figure 6a and b shows the magnetic hysteresis loops at 5 K for FeNi-PEG and

FeNi-PVP, respectively. The hysteresis loop reveals that the samples have the symmetric hystereis loop behavior of ferromagnetic materials. Values of saturation magnetization ($M_s$), remanence-to-saturation magnetization ratio ($M_r/M_s$), and the coercivity ($H_c$) measured in ZFC and FC conditions at 5K are listed in Table 1 and 2. These values are in good agreement with the earlier reported values on FeNi-PMMA nanocomposite system [18]. The decrease in the Ms values with the increase in MW of polymers is obvious due to the overall increase in the sample weight. The $H_c$ values of FeNi-PVP samples are found to be higher than that of FeNi-PEG samples. The presence of relatively ordered FeNi nanocrystallites with higher magnetocrystalline anisotropy may contribute to the high covercivity in FeNi-PVP compared to FeNi-PEG nanocomposites. The slow reaction rate in the presence of PVP may provide enough time for the formation of ordered structure of FeNi nanocrystallites in the composite. Furthermore, in the framework of Stoner-Wohlfarth model [30] in the ideal case of well-separated fine particles $M_r/M_s$ ratio is 0.5. Thus, the fact that $M_r(0)/M_s(0)$ values are below 0.5 (Table 1 & 2) indicate relatively low interparticle distances between FeNi nanocrystallites in all samples.

The hysteresis loop shift and enhanced Mr/Ms ratio under the field cooled condition indicate the presence of exchange bias effect in the FeNi nanocomposites, insets of Fig. 6a and b. An exchange bias field measured at 5 K of around 50 Oe and 77 Oe has been observed for FeNi-PEG and FeNi-PVP samples, respectively. The presence of exchange bias in these nanocomposites may arise from the surface ferrimagnetic spins or spin glass and ferromagnetic FeNi core spin interactions [31].

In summary, spherical well-dispersed FeNi-polymer nanocomposites were prepared using chemical reduction method in the presence of PVP and PEG polymers. The as synthesized nanocomposite particles were formed by the aggregate of finer FeNi nanocrystallites embedded in polymer matrix. Although the size of the nanocomposite particles depend on the type of polymers used, the observed magnetic properties of the nanocomposites are found to be independent of the type of polymers. Thus, present synthesis technique offers unique route to preparing magnetic nanocoposites whose magnetic properties are independent of the size of particles. The controlled morphology, the narrow size distribution, sub-micron size range, and with biocompatible matrix PVP/PEG, could have applications in the field of biomedical engineering, for example hyperthermia application or cell/drug separation. These particles could also find application in microwave devices as electromagnetic-wave absorbing materials [32].

**Table Captions:**

**Table 1:** ZFC magnetic parameters for FeNi-polymer nanocomposites measured at 5K.

**Table 2:** FC magnetic parameters for FeNi-polymer nanocomposites measured at 5K.

**Figure Captions:**

**Figure 1a**: XRD pattern of FeNi-PEG nanocomposite samples.

**Figure 1b**: XRD pattern of FeNi-PVP nanocomposite samples.

**Figure 2a**: FeNi-PEG (a) MW 200, (b) MW 1000, (c) MW 8000, and (d) MW 20000. The scale bar length is equivalent to 100 nm.

**Figure 2b**: FeNi-PVP (a) MW 10000, (b) MW 29000, (c) MW 40000, and (d) MW 55000. The scale bar length is equivalent to 100 nm.

**Figure 3**: Unheated FeNi-polymer nanocomposite particles size, (O) FeNi-PEG and (Δ) FeNi-PVP, as a function of MW of polymers. Inset shows the dependence of FeNi nanocrystallite size on MW of polymers. Solid symbols are for samples heated at 500 $^o$C.

**Figure 4**: (I) DTA curves measured in air of (a) Pure PEG MW 8000, (b) Pure PVP MW 10,000, (c) FeNi-PEG (MW8000), and (d) FeNi-PVP (MW 10,000). Inset of (I) shows XRD of samples heated at 500 $^o$C in air. (II) TGA curves measured in air of (a) Pure PEG MW 8000, (b) Pure PVP MW 10,000, (c) FeNi-PEG (MW8000), and (d) FeNi-PVP (MW 10,000).

**Figure 5**: Magnetic susceptibility as a function of temperature at 1 kOe for composite samples. Open and solid symbols represent ZFC and FC conditions.

**Figure 6a:** Hysteresis loops of FeNi-PEG nanocomposite samples measured at 5K. Inset shows close view of ZFC and FC hysteresis loops of FeNi-PEG (MW 20,000) measured at 5K.

**Figure 6b:** Hysteresis loops of FeNi-PVP nanocomposite samples measured at 5K. Inset shows close view of ZFC (open symbol) and FC (close symbol) hysteresis loops of FeNi-PVP (MW 55,000) measured at 5K.

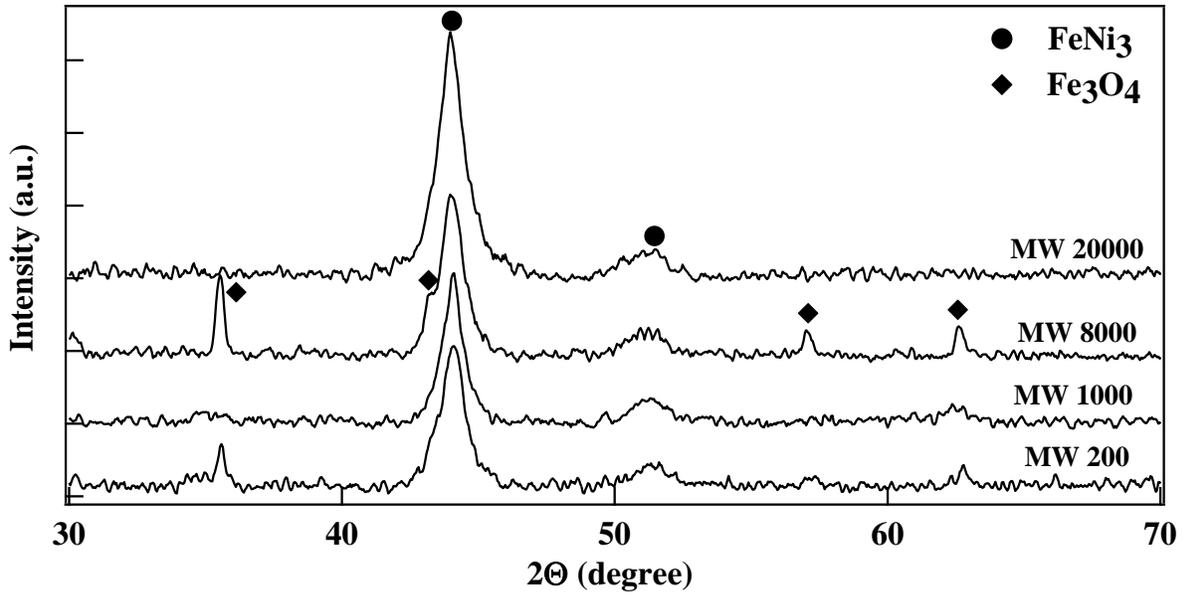

**Figure 1a**: XRD pattern of FeNi-PEG nanocomposite samples.

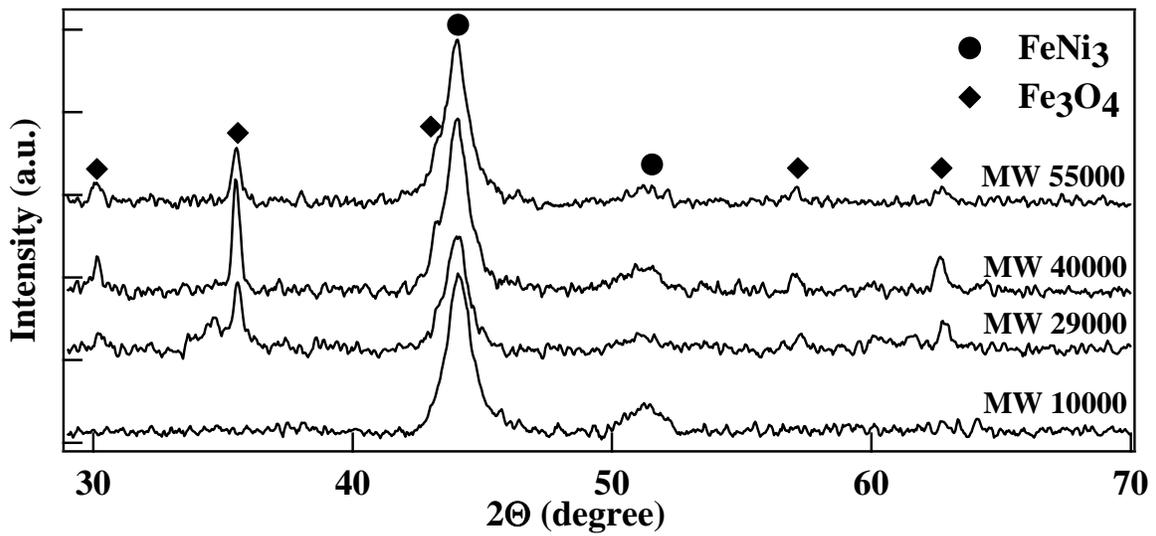

**Figure 1b**: XRD pattern of FeNi-PVP nanocomposite samples.

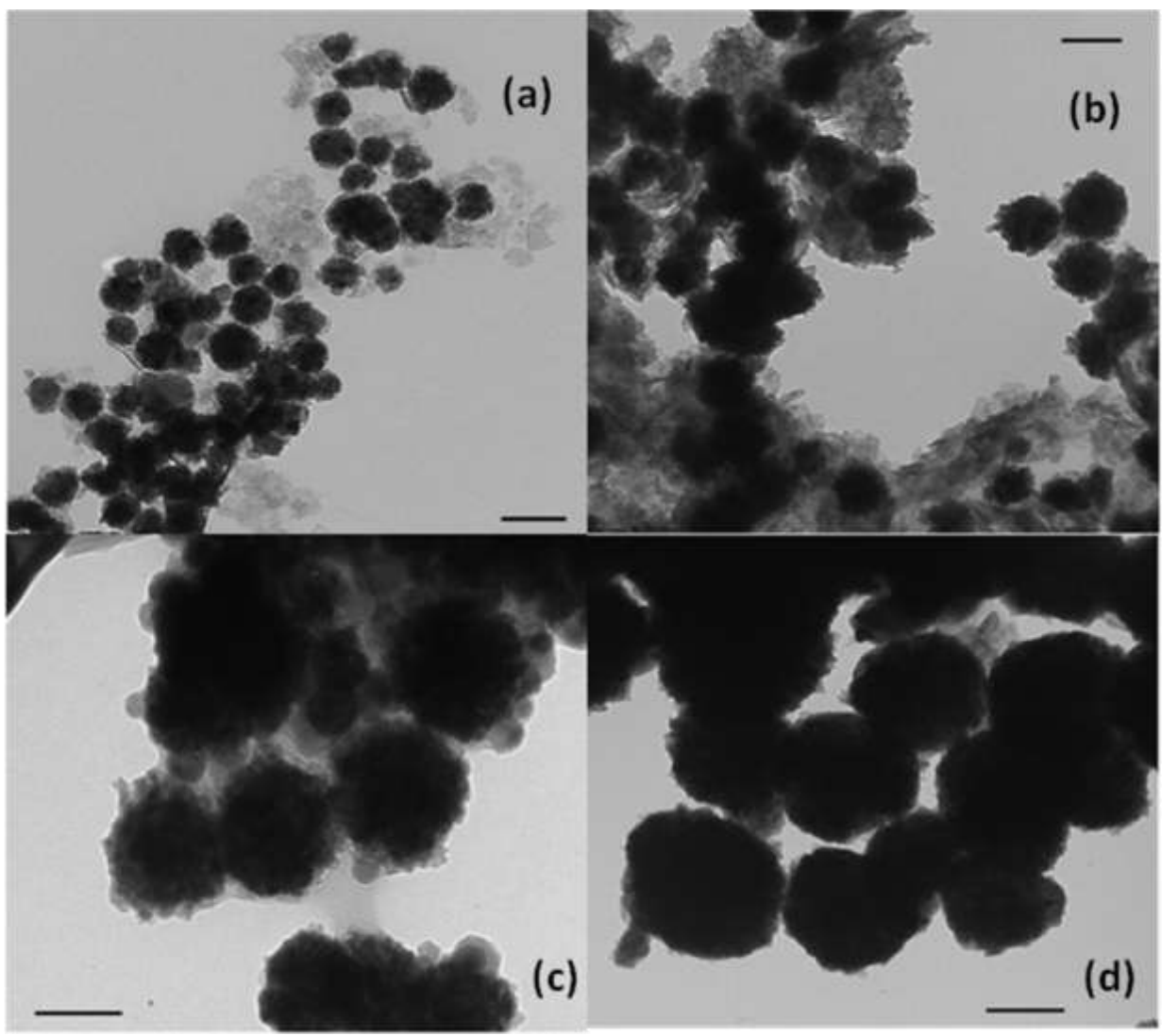

**Figure 2a**: FeNi-PEG (a) MW 200, (b) MW 1000, (c) MW 8000, and (d) MW 20000. The scale bar length is equivalent to 100 nm.

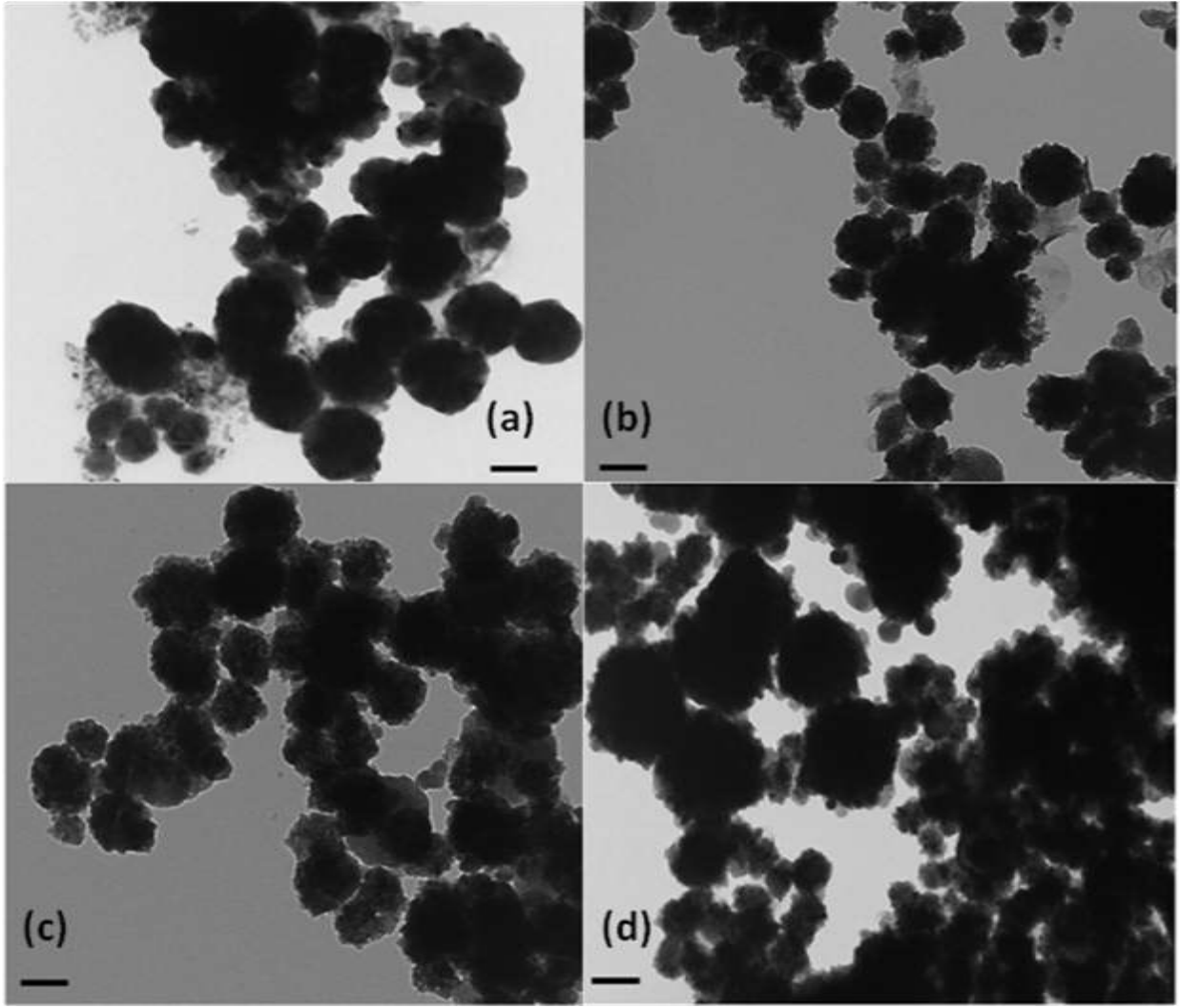

**Figure 2b**: FeNi-PVP (a) MW 10000, (b) MW 29000, (c) MW 40000, and (d) MW 55000. The scale bar length is equivalent to 100 nm.

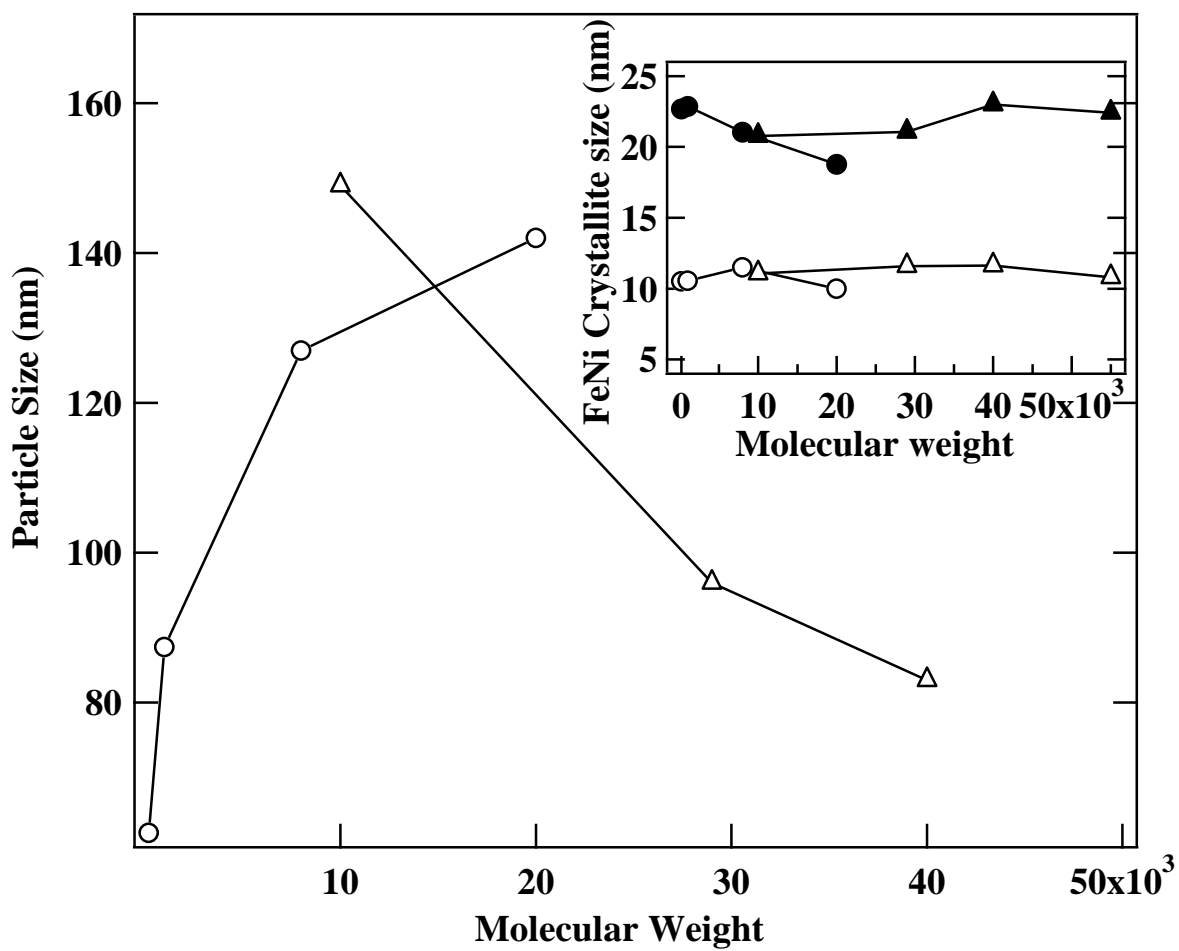

**Figure 3**: Unheated FeNi-polymer nanocomposite particles size, (O) FeNi-PEG and (Δ) FeNi-PVP, as a function of MW of polymers. Inset shows the dependence of FeNi nanocrystallite size on MW of polymers. Solid symbols are for samples heated at 500 °C.

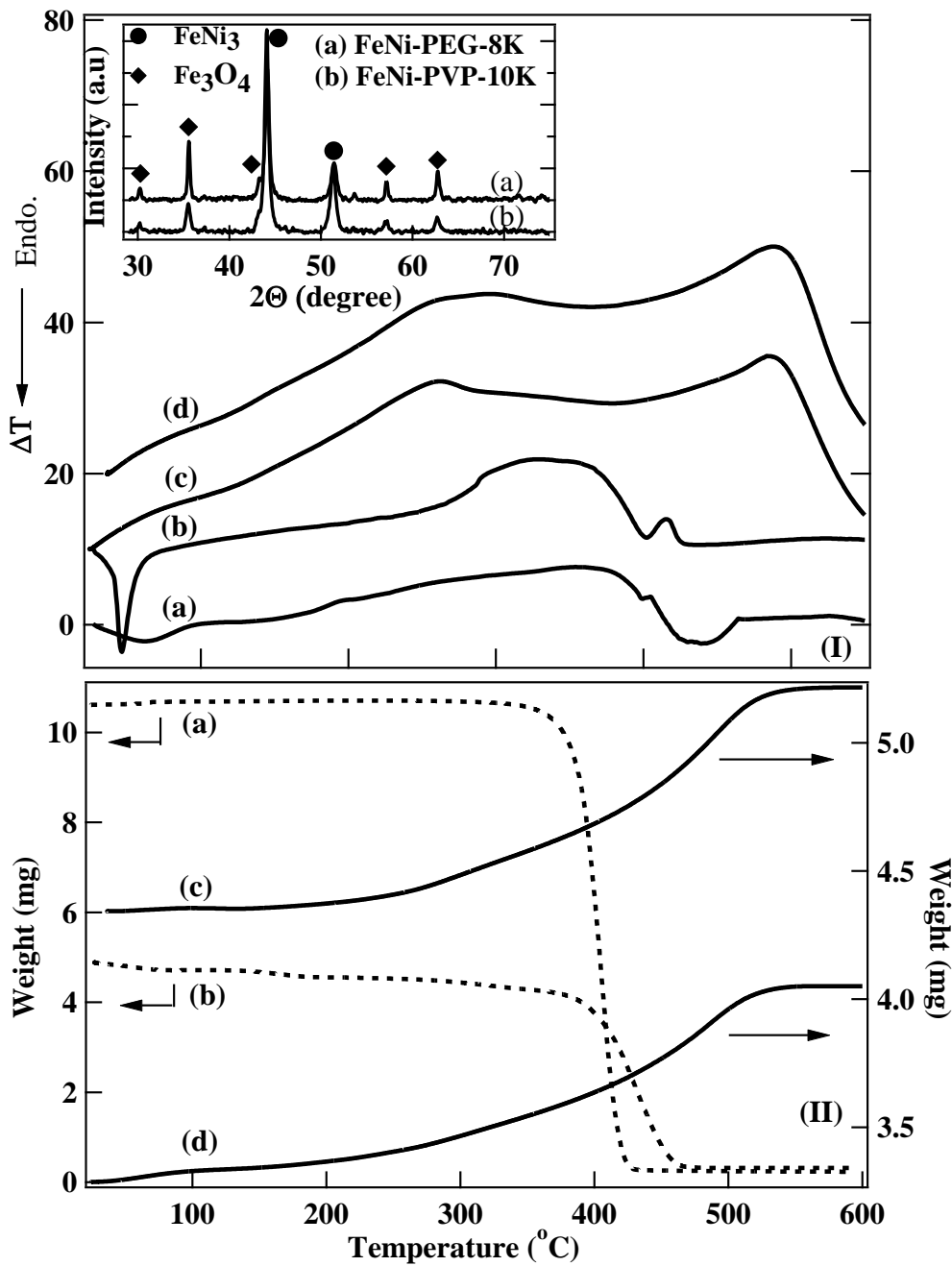

**Figure 4**: (I) DTA curves measured in air of (a) Pure PEG MW 8000, (b) Pure PVP MW 10,000, (c) FeNi-PEG (MW8000), and (d) FeNi-PVP (MW 10,000). Inset of (I) shows XRD of samples heated at 500 °C in air. (II) TGA curves measured in air of (a) Pure PEG MW 8000, (b) Pure PVP MW 10,000, (c) FeNi-PEG (MW8000), and (d) FeNi-PVP (MW 10,000).

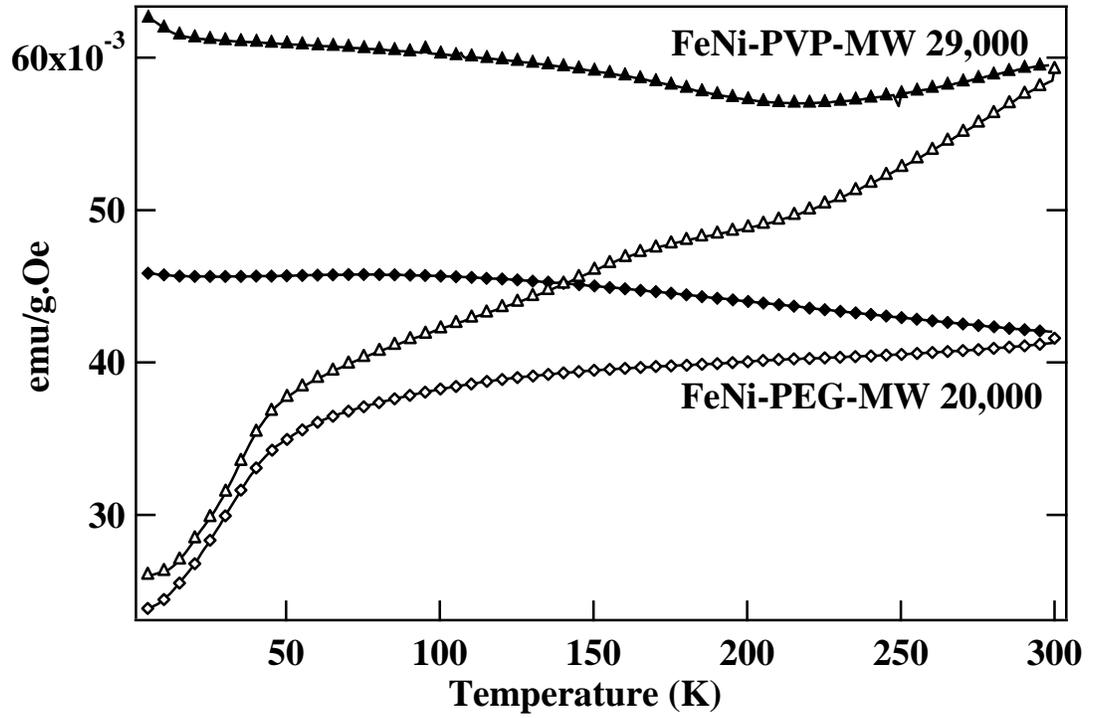

**Figure 5:** Magnetic susceptibility as a function of temperature at 1 kOe for composite samples. Open and solid symbols represent ZFC and FC conditions, respectively.

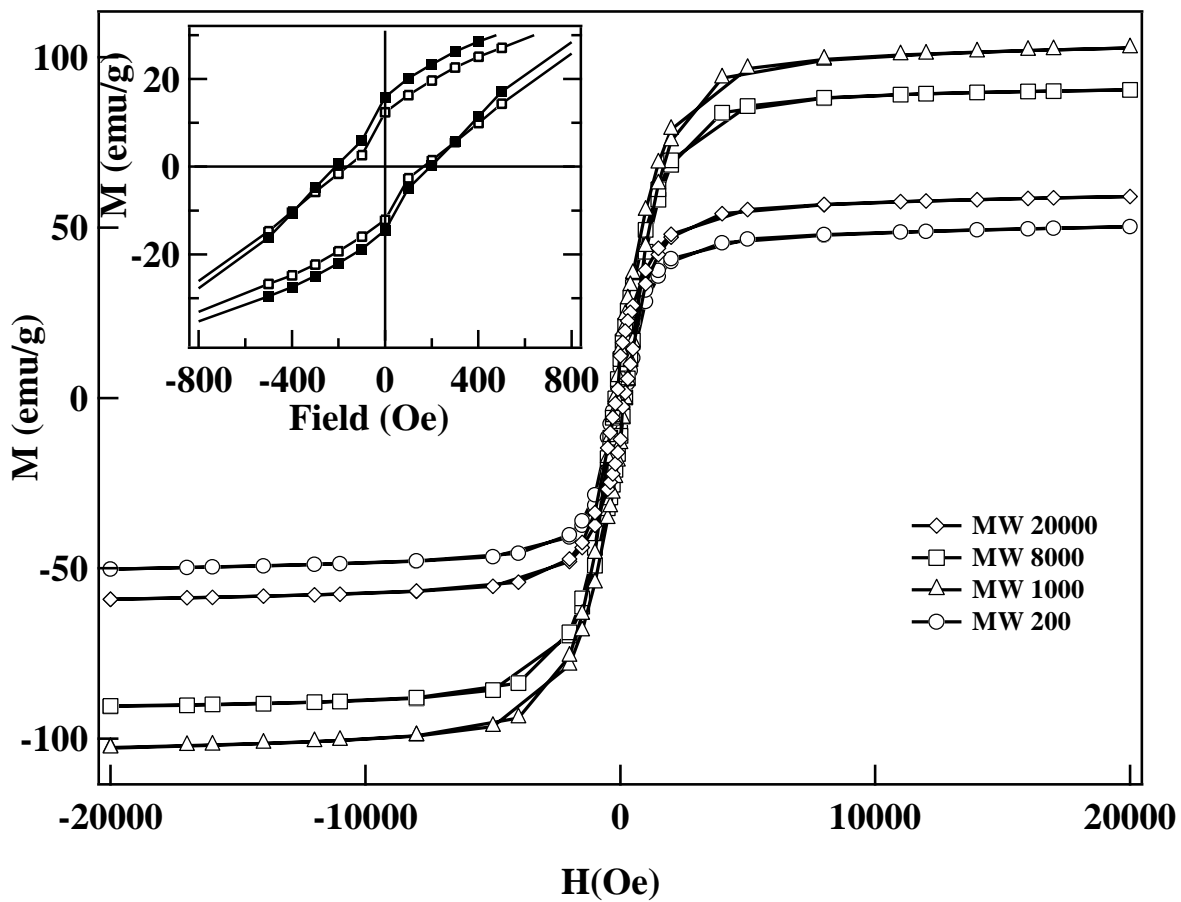

**Figure 6a:** Hysteresis loops of FeNi-PEG nanocomposite samples measured at 5K. Inset shows close view of ZFC (open symbol) and FC (close symbol) hysteresis loops of eNi-PEG (MW 20,000) measured at 5K.

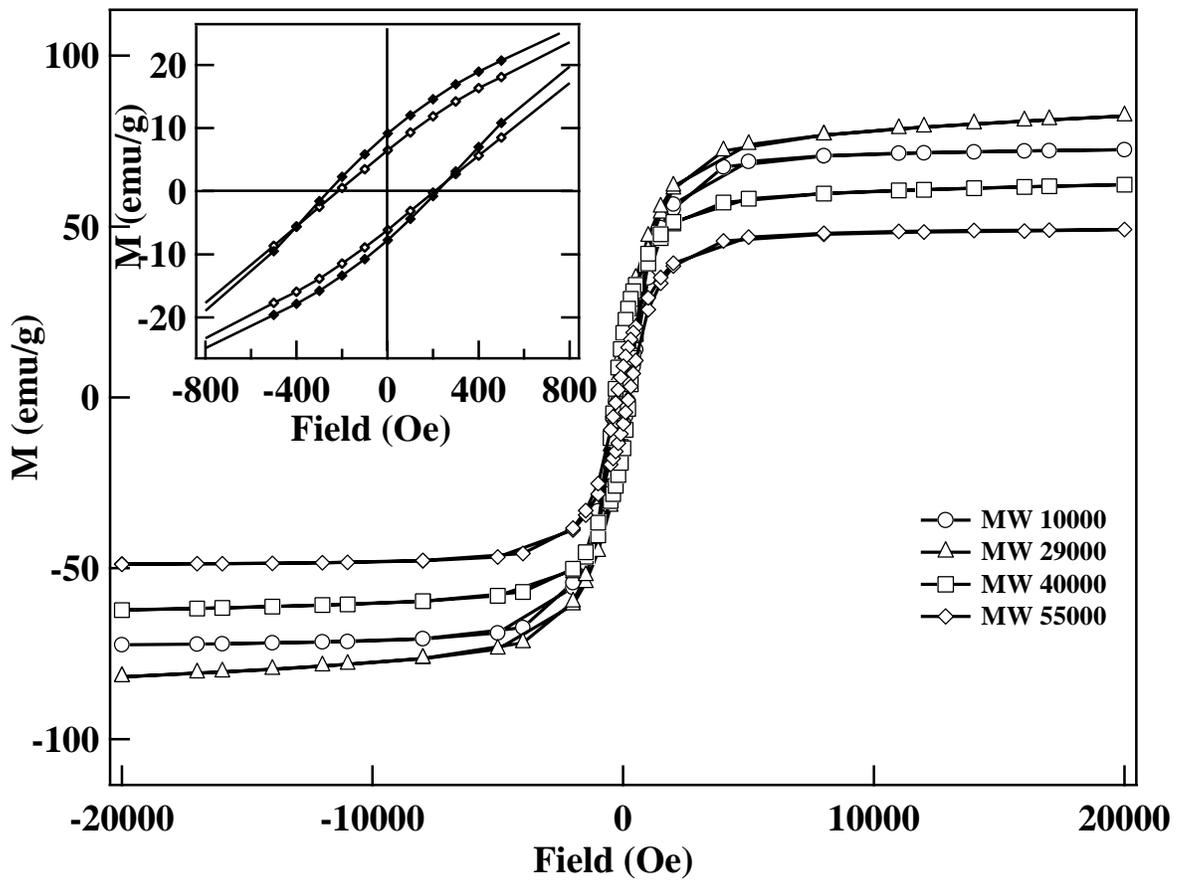

**Figure 6b:** Hysteresis loops of FeNi-PVP nanocomposite samples meaured at 5K. Inset shows close view of ZFC (open symbol) and FC (close symbol) hysteresis loops of FeNi-PVP (MW 55,000) measured at 5K.

**Table 1**: ZFC magnetic parameters for FeNi-polymer nanocomposites measured at 5K.

| MW PEG | PVP | Mr (emu/g) | Ms (emu/g) | Mr/Ms | Hc (Oe) |
|---|---|---|---|---|---|
| 200 | | 9 | 50 | 0.17 | 186 |
| 1000 | | 14 | 102 | 0.13 | 205 |
| 8000 | | 11 | 90 | 0.12 | 199 |
| 20000 | | 12 | 82 | 0.14 | 153.5 |
| | 10000 | 9 | 72 | 0.12 | 218 |
| | 29000 | 11 | 80 | 0.13 | 227 |
| | 40000 | 12 | 62 | 0.19 | 260 |
| | 55000 | 6 | 48 | 0.13 | 213 |

**Table 2**: FC magnetic parameters for FeNi-polymer nanocomposites measured at 5K.

| MW PEG | PVP | Mr (emu/g) | Ms (emu/g) | Mr/Ms | Hc (Oe) | $H_E$ (Oe) |
|---|---|---|---|---|---|---|
| 200 | | 12 | 50 | 0.24 | 213 | 57 |
| 1000 | | 17 | 102 | 0.16 | 242 | 55 |
| 8000 | | 16 | 91 | 0.17 | 227 | 57 |
| 20000 | | 20 | 84 | 0.23 | 272 | 57 |
| | 10000 | 11 | 72 | 0.15 | 251 | 42 |
| | 29000 | 15 | 82 | 0.18 | 251 | 72 |
| | 40000 | 17 | 62 | 0.27 | 292 | 75 |
| | 55000 | 9 | 49 | 0.18 | 240 | 45 |